\documentclass[runningheads]{CMSIM}

\usepackage{graphicx} 

\usepackage{amsmath,amssymb}

\renewcommand{\thefootnote}

\title*{Dirac Theory as a Relativistic Flow}

\titlerunning{\it Dirac Theory as Relativistic Flow}

\author{
Asher Yahalom
}

\authorrunning{\it Asher Yahalom\inst{1}}

\institute{
Ariel University, Ariel 40700, Israel\\
(E-mail: {\tt asya@ariel.ac.il})
}

\begin{document}
\thispagestyle{empty}
\maketitle
\setlength{\leftskip}{0pt}
\setlength{\headsep}{16pt}
\footnote{
\begin{tabular}{p{11.2cm}r}
\small {\it $17^{th}$CHAOS Conference Proceedings, 11 - 14 June 2024, Chania, Crete, Greece} \\
\small C. H. Skiadas (Ed)\\
\small \textcopyright {} 2024 ISAST
 \end{tabular}
 }
\begin{abstract}
In previous papers we have shown how Schr\"{o}dinger's equation which includes an electromagnetic field interaction can be deduced from a fluid dynamical Lagrangian of a charged potential flow that interacts with an electromagnetic field. The quantum behaviour was derived from Fisher information terms which were added to the classical Lagrangian. It was thus shown that a quantum mechanical system is drived by information and not only electromagnetic fields.

This program was applied also to Pauli's equations by removing the restriction of potential flow and using the Clebsch formalism. Although the analysis was quite successful there were still terms that did not admit interpretation, some of them can be easily traced to the relativistic Dirac theory. Here we repeat the analysis for a relativistic flow, pointing to a new approach for deriving relativistic quantum mechanics.
\end{abstract}

\newcommand{\beq} {\begin{equation}}
\newcommand{\enq} {\end{equation}}
\newcommand{\ber} {\begin {eqnarray}}
\newcommand{\enr} {\end {eqnarray}}
\newcommand{\eq} {equation}
\newcommand{\eqn} {equation }
\newcommand{\eqs} {equations }
\newcommand{\ens} {equations}
\newcommand {\er}[1] {equation (\ref{#1}) }
\newcommand {\ern}[1] {equation (\ref{#1})}
\newcommand {\ers}[1] {equations (\ref{#1})}
\newcommand {\Er}[1] {Equation (\ref{#1}) }
\newcommand{\br} {\bar{r}}
\newcommand{\tnu} {\tilde{\nu}}
\newcommand{\rhm}  {{\rho \mu}}
\newcommand{\sr}  {{\sigma \rho}}
\newcommand{\bh}  {{\bar h}}
\newcommand {\Sc} {Schr\"{o}dinger}
\newcommand {\SE} {Schr\"{o}dinger equation }
\newcommand {\bR} {{\bf R}}
\newcommand {\bX} {{\bf X}}
\newcommand{\ce}  {continuity equation }
\newcommand{\ces} {continuity equations }
\newcommand{\hje} {Hamilton-Jacobi equation }
\newcommand{\hjes} {Hamilton-Jacobi equations }
\newcommand{\bp}  {\bar{\psi}}
\newcommand{\va}  {{\vec \alpha}}
\newcommand{\mn}  {{\mu \nu}}

\section {Introduction}

Quantum mechanics is commonly understood through the Copenhagen interpretation, which views the quantum wave function as an epistemological tool, used solely for predicting measurement probabilities, aligning with Kantian philosophy that denies human ability to understand things "as they are" (ontology) \cite{Kant}. However, another prominent school of thought interprets quantum mechanics differently, believing in the wave function's reality. This view, supported by Einstein and Bohm \cite{Bohm,Holland,DuTe}, considers the wave function as a real entity akin to an electromagnetic field. This perspective led to alternative interpretations, such as Madelung's fluid realization \cite{Madelung,Complex}, which proposes that the square of the wave function's modulus represents a fluid density, and its phase represents the fluid's velocity potential. However, this approach is limited to spinless electron wave functions and cannot account for a full set of attributes even for slow-moving electrons.

Wolfgang Pauli \cite{Pauli} introduced a non-relativistic quantum equation for a spinor in 1927. This equation utilizes a two-dimensional operator matrix Hamiltonian. It was shown that such a theory can be interpreted through fluid dynamics \cite{Spflu}. This interpretation is significant because proponents of the Copenhagen interpretation of quantum mechanics often use the concept of spin as evidence that nature has inherently quantum elements without classical analogues or interpretations.

Holland \cite{Holland} and others provided a Bohmian analysis of the Pauli equation, but they did not address the analogy between Pauli theory and fluid dynamics or the notion of spin vorticity. Thus, spin fluid dynamics for a single electron with spin \cite{Spflu} was introduced subsequently. 

Interpreting Pauli's spinor in terms of fluid density and velocity variables connects us to the $19^{th}$-century work of Clebsch, which is closely tied to the Eulerian variational analysis of fluids. Clebsch \cite{Clebsch1,Clebsch2} and, much later, Davidov \cite{Davidov} described variational principles for barotropic fluid dynamics. Clebsch introduced a four-function variational principle for an Eulerian barotropic fluid, and Davidov aimed to quantize fluid dynamics, though his work was not well-known in the West due to being written in Russian. Eckart \cite{Eckart} provided a variational description for Lagrangian fluid dynamics, which differs from the variational approach to Eulerian fluid dynamics.

Initial attempts to formulate Eulerian fluid dynamics using variational principles in the English literature were made by Herivel  \cite{Herivel}, Serrin \cite{Serrin}, and Lin \cite{Lin}. However, their methods were complicated, relying on numerous Lagrange multipliers and auxiliary potentials, involving between eleven to seven independent functions—more than the four required for the Eulerian and continuity equations of barotropic flow, making these methods impractical.

Seliger and Whitham \cite{Seliger} reintroduced Clebsch's variational formalism using only four variables for barotropic flow. Lynden-Bell and Katz \cite{LynanKatz} proposed a variational principle in terms of two functions, load \( \lambda \) and density \( \rho \), but their approach had an implicit definition of velocity \( \vec{v} \), requiring the solution of a partial differential equation to determine \( \vec{v} \) in terms of \( \rho \) and \( \lambda \) and its variations. Yahalom and Lynden-Bell \cite{YahLyndeb} addressed this limitation by adding one more variational variable, allowing for arbitrary (unconstrained) variations and providing an explicit definition of \( \vec{v} \).

A key challenge in interpreting quantum mechanics through fluid dynamics lies in understanding thermodynamic quantities. In traditional fluids, concepts like specific enthalpy, pressure, and temperature relate to specific internal energy, which is a unique function of entropy and density defined by the equation of state. This internal energy can be explained through the microscopic composition of the fluid using statistical physics, based on the interactions of atoms, ions, electrons and molecules through electromagnetic fields.

However, a quantum fluid lacks such structure. Yet, equations for both spinless \cite{Madelung,Complex} and spin \cite{Spflu} quantum fluid dynamics show terms analogous to internal energies. This raises the question: where do these internal energies come from? Suggesting that quantum fluids have a microscopic substructure contradicts empirical evidence that electrons are point particles.

The answer lies in measurement theory, particularly Fisher information \cite{Fisher,Fisherspin}, a measure of the quality of any quantity's measurement. It has been shown \cite{Fisherspin} that Fisher information corresponds to the internal energy of a non-relativistic spinless electron (up to a proportionality constant)and can partly explain the internal energy of non-relativistic electron with spin \cite{Fisherspin}. 

There has been an attempt to derive most physical theories from Fisher information by Frieden \cite{Frieden}, however, in this approach there is always a $J$ component to the Lagrangian (in addition to the Fisher information) which is unique to each physical system and is chosen without justification such that the desired Lagrangian is derived.

At the time of Clebsch, relativity was not introduced yet hence there was no need to write a variational principle for an Eulerian relativistic flow (which invariant under a Lorentz transformation). This was recently rectified in a series of papers \cite{FluRel,FluRel2,FluRel3} in which relativistic Clebsch fluid dynamics was introduced. It was shown also that relativistic Clebsch fluid dynamics can lead to  relativistic quantum mechanics by adding a Lorentz invariant Fisher information term. For null vorticity and low velocities this variational principle reduces to the Schr\"{o}dinger variational principle. 

Thus it is now needed to compare the fluid derived relativistic quantum mechanics to the more prevalent Dirac theory which is the current established theory of relativistic quantum mechanics.
This comparison involves several steps the first of which is to express the theory in terms of four variables of a fluid (velocity vector \& density) rather than eight (a complex four spinor) of Dirac. This will be done in the following sections.

\section{Dirac Theory}

The theory of Dirac is defined in terms of the equation (we initially neglect the electromagnetic interactions):
\beq
\left(i\hbar \gamma^\mu \partial_\mu - m c\right) \Psi = 0
\label{Diracequa}
\enq
$\Psi$ is a four dimensional complex column vector (spinor). And $\gamma^\mu$ are four dimensional complex matrices  satisfying the anticommutation relations:
\beq
\left\{\gamma^\mu,\gamma^\nu\right\} = 2 \eta^{\mu \nu} I_4, \qquad
\eta^{\mu \nu} ={\rm diag} (+1,-1,-1,-1)
\label{gammanticom}
\enq
$I_4$ is a unit matrix in four dimensions. In what follows Greek indices: $\mu, \nu \in \{0,1,2,3\}$
and Latin indices: $i,j,k \in \{1,2,3\}$. There are multiple representations of $\gamma^\nu$, we
shall use the following representation:
\beq
\gamma^0= \left( \begin{array}{cc} I_2 & 0 \\ 0 & -I_2 \end{array} \right) \qquad
\gamma^i= \left( \begin{array}{cc} 0 & \sigma^i \\ -\sigma^i & 0 \end{array} \right)
\label{gammadef}
\enq
\beq
\sigma^{1} = \left( \begin{array}{cc} 0 & 1 \\ 1 & 0 \end{array} \right), \qquad
\sigma^{2} = \left( \begin{array}{cc} 0 & -i \\ i & 0 \end{array} \right), \qquad
\sigma^{3} = \left( \begin{array}{cc} 1 & 0 \\ 0 & -1 \end{array} \right).
\label{sigma}
\enq
and $I_2$ is a unit matrix in two dimensions. \Er{Diracequa} may be integrated provided
initial conditions are supplied, that is:
\beq
\Psi(0,\vec x) = \Psi_0(\vec x)
\label{Psi0}
\enq
The theory seems disconnected from any fluid dynamic interpretation as $\Psi$ depends on eight scalar quantities while
a barotropic fluid theory depends on only four variables (half the required amount).
However, the theory can be expressed in terms of less variables as follows. First we
write the four dimensional spinor in terms of two dimensional spinors:
\beq
\Psi = \left( \begin{array}{c} \psi_1 \\ \psi_2\end{array} \right).
\label{spin2D}
\enq
This form induces by \ern{Psi0} initial condition on both $\psi_1$ and $\psi_2$:
\beq
\psi_1(0,\vec x) = \psi_{10}(\vec x), \quad \psi_2(0,\vec x) = \psi_{20}(\vec x), \qquad
\Psi_0 = \left( \begin{array}{c} \psi_{10} \\ \psi_{20} \end{array} \right).
\label{psi120}
\enq
Substituting \ern{spin2D} in \ern{Diracequa} we obtain:
\beq
(i \hbar \partial_0 - mc) \psi_1 + i \hbar \sigma^i \partial_i  \psi_2 = 0, \qquad
(i \hbar \partial_0 + mc) \psi_2 + i \hbar \sigma^i \partial_i  \psi_1 = 0
\label{spin2Deq}
\enq
Introducing the hatted variables:
\beq
\hat \psi_1 \equiv  e^{-\frac{imc}{\hbar}x_0}\psi_1 = e^{-\frac{imc^2}{\hbar}t}\psi_1, \qquad
\hat \psi_2 \equiv e^{-\frac{imc}{\hbar}x_0}\psi_2 = e^{-\frac{imc^2}{\hbar}t}\psi_2.
\label{hatpsi}
\enq
We can substitute:
\beq
 \psi_1 = e^{+\frac{imc}{\hbar}x_0} \hat \psi_1 , \qquad
 \psi_2 = e^{+\frac{imc}{\hbar}x_0} \hat \psi_2 .
\label{hatpsi2}
\enq
in \ern{spin2Deq} and obtain the simplified set of equations:
\beq
(i \hbar \partial_0 - 2mc) \hat \psi_1 + i \hbar \sigma^i \partial_i  \hat \psi_2 = 0, \qquad
i \hbar \partial_0  \hat \psi_2 + i \hbar \sigma^i \partial_i \hat \psi_1 = 0
\label{spin2Deq2}
\enq
The initial conditions for this equations at $x_0=0$ are the same as before, because
$\psi$ and $\hat \psi$ are the same at that particular time:
 \beq
\hat \psi_1(0,\vec x) = \psi_{10}(\vec x), \quad \hat \psi_2(0,\vec x) = \psi_{20}(\vec x).
\label{psih120}
\enq
The second equation for $\hat \psi_2$ can be readily solved if $\hat \psi_1$ is known:
\beq
\hat \psi_2 (x_0,\vec x)[\hat \psi_1] = \hat \psi_2 (0,\vec x) - \sigma^i \partial_i \int_{0}^{x_0}\hat \psi_1 (x'_0,\vec x) d x'_0
\label{hatpsi2sol}
\enq
Introducing the auxiliary variable:
\beq
int\hat \psi_1 \equiv \int_{0}^{x_0}\hat \psi_1 (x'_0,\vec x) d x'_0
\Rightarrow
\partial_0 int\hat \psi_1 = \hat \psi_1,
\partial^2_0 int\hat \psi_1 = \partial_0 \hat \psi_1
\label{inthatpsi1}
\enq
and the time independent spinor:
\beq
W(\vec x) \equiv - \sigma^k \partial_k \hat \psi_2 (0,\vec x),
\label{Wdef}
\enq
we may write \ern{spin2Deq2} and thus the Dirac theory in the form:
\beq
(\partial^\mu \partial_\mu + 2 \frac{imc}{\hbar}\partial_0) int\hat \psi_1 = W (\vec x), \qquad
\hat \psi_2 (x_0,\vec x) = \hat \psi_2 (0,\vec x) - \sigma^i \partial_i int\hat \psi_1
\label{spin2Deq3}
\enq
Hence the mathematical problem of Dirac theory is to solve the first part of \ern{spin2Deq3} because the second equation is just a relation giving us $\hat \psi_2$ explicitly in terms of $\hat \psi_1$. Moreover, if we have a solution
for $\hat \psi_1$ then $int\hat \psi_1$ follows immediately from \ern{inthatpsi1}. Taking the temporal partial derivative of the first equation in (\ref{spin2Deq3}) it follows that $\hat \psi_1$
must satisfy the equation:
\beq
(\partial^\mu \partial_\mu + 2 \frac{imc}{\hbar}\partial_0) \hat \psi_1 = 0
\label{spin2Deq4}
\enq
The initial conditions of this second order equation are fixed by the initial conditions of
 \ern{spin2Deq2} because those conditions also fix the first derivative in time $x_0=0$.
\beq
(i \hbar \partial_0 - 2mc) \hat \psi_1|_{x_0=0} + i \hbar \sigma^i \partial_i  \psi_{20} = 0, \qquad
\partial_0  \hat \psi_2|_{x_0=0} +  \sigma^i \partial_i \psi_{10} = 0.
\label{initialcond}
\enq
Or more simply as:
\beq
\partial_0 \hat \psi_1|_{x_0=0} = \frac{2mc}{i\hbar} \psi_{10} - \sigma^i \partial_i \psi_{20}
\label{initialcond2}
\enq
 As we are given both the initial condition of the function and the initial condition of its first derivative, the solution of the second order differential \ern{spin2Deq4} are fixed, and its solution is the entire content of the Dirac theory. We notice at this point that one can reintroduce the original function $\psi_1$ using \ern{hatpsi} which will result in the Klein Gordon
 equation:
\beq
(\partial^\mu \partial_\mu + \frac{m^2 c^2}{\hbar^2})  \psi_1 = 0
\label{kleingordon}
\enq
with the initial conditions:
\beq
\psi_1|_{x_0=0} =  \psi_{10}, \qquad
\partial_0 \psi_1|_{x_0=0} = \frac{mc}{i\hbar}  \psi_{10} - \sigma^i \partial_i  \psi_{20}.
\label{initialcond3}
\enq
which is also equivalent to Dirac's theory. However, notice first that in this case the Klein Gordon equation is an equation for a two dimensional spinor nor a scalar or even a complex scalar. Second,
the physical interpretation in Dirac theory is quite different with respect to the original Klein Gordon theory. In particular the conserved probability four current is:
\beq
J^\mu \equiv \bar{\Psi}\gamma^\mu \Psi, \qquad \bar{\Psi} \equiv \Psi^\dagger \gamma^0
\label{Jprob}
\enq
Thus we obtain the probability density:
\beq
J^0 = \bar{\Psi}\gamma^0\Psi = \Psi^\dagger (\gamma^0)^2 \Psi = \Psi^\dagger \Psi
= \psi_1^\dagger \psi_1 + \psi_2^\dagger \psi_2 \ge 0.
\label{Jprob2}
\enq
This is quite different from $J^0$ in the original Klein Gordon theory which could become negative, and thus unphysical. Nevertheless, we have shown that from a mathematical point of view both theories have identical equations but different mathematical dependent variables. In the Klein Gordon theory we consider (complex) scalars and in the Dirac theory we consider spinors.
We are in a better position now to show the analogies with relativistic flows as at least both theories depend on identical number of dependent variables that is four scalar functions.

\section{Variational description}

\Er{kleingordon} can be deduced from a variational principle using the Lagrangian density:
\beq
{\cal L}_{KG} \equiv  m \left( \frac{\hbar^2}{m^2} \partial^\mu  \psi_1^\dagger  \partial_\mu \psi_1  - c^2 \psi_1^\dagger \psi_1\right), \qquad {\cal A}_{KG} \equiv \int d^4 x {\cal L}_{KG}
\label{LKG}
\enq
provided that the variations are constrained in a suitable manner on the spatial and temporal boundaries. This is not the traditional Dirac Lagrangian density but has the same mathematical content non the less, as we have shown in the previous section. Let us write the two dimensional spinor $\psi_1$ in terms of its up and down components:
\beq
\psi_1 = \left( \begin{array}{c} \psi_\uparrow \\ \psi_\downarrow\end{array} \right).
\label{psiupdo}
\enq
Inserting \ern{psiupdo} into \ern{LKG} we obtain:
\beq
{\cal L}_{KG} =   m \left( \frac{\hbar^2}{m^2} \partial^\mu  \psi_\uparrow^*  \partial_\mu \psi_\uparrow
- c^2 \psi_\uparrow^* \psi_\uparrow
+ \frac{\hbar^2}{m^2} \partial^\mu  \psi_\downarrow^*  \partial_\mu \psi_\downarrow
- c^2 \psi_\downarrow^* \psi_\downarrow \right).
\label{LKG2}
\enq
We now write the up and down wave functions in an amplitude and phase representation:
\beq
\psi_\uparrow = R_\uparrow e^{i \frac{m}{\hbar} \nu_\uparrow}, \qquad
\psi_\downarrow = R_\uparrow e^{i \frac{m}{\hbar} \nu_\downarrow}
\label{phaamp}
\enq
Substituting \ern{phaamp} into \ern{LKG2} will lead to the form:
\ber
{\cal L}_{KG} &=& {\cal L}_{KGq} + {\cal L}_{KGc}
\nonumber \\
{\cal L}_{KGq} &\equiv& \frac{\hbar^2}{m} \left(\partial^\mu  R_\uparrow  \partial_\mu
R_\uparrow +\partial^\mu  R_\downarrow  \partial_\mu R_\downarrow\right)
\nonumber \\
{\cal L}_{KGc} &\equiv& m \left(R_\uparrow^2\left(\partial^\mu  \nu_\uparrow  \partial_\mu \nu_\uparrow - c^2\right)+ R_\downarrow^2\left(\partial^\mu  \nu_\downarrow  \partial_\mu \nu_\downarrow - c^2\right)\right).
\label{LKG3}
\enr
in which we have partitioned ${\cal L}_{KG}$ into a quantum part ${\cal L}_{KGq}$ and a classical part ${\cal L}_{KGc}$. In the classical limit in which $\hbar \rightarrow 0$:
\beq
\lim_{\hbar \rightarrow 0} {\cal L}_{KGq} = 0 \quad \Rightarrow \quad
\lim_{\hbar \rightarrow 0}{\cal L}_{KG} = {\cal L}_{KGc}
\label{LKGnonql}
\enq
We introduce a mass density and an angle $\theta$ in the following natural
way:
\beq
\bar \rho \equiv m \left( R_\uparrow^2 + R_\downarrow^2 \right), \quad
\tan \theta \equiv \frac{R_\downarrow}{R_\uparrow}, \quad \Rightarrow \quad
R_\uparrow = \sqrt{\frac{\bar \rho}{m}} \cos \theta, \quad R_\downarrow = \sqrt{\frac{\bar \rho}{m}} \sin \theta
\label{barho}
\enq
It thus follows that:
\beq
{\cal L}_{KGc} = \bar \rho \left[\cos^2 \theta \partial^\mu  \nu_\uparrow  \partial_\mu \nu_\uparrow +\sin^2 \theta \partial^\mu  \nu_\downarrow  \partial_\mu \nu_\downarrow - c^2\right].
\label{LKGc}
\enq
Now let us set:
\ber
\nu &\equiv& \nu_\uparrow, \qquad \beta \equiv \nu_\downarrow - \nu_\uparrow,
\nonumber \\
\alpha &\equiv& \frac{-\partial_\mu \nu \partial^\mu \beta
\pm \sqrt{(\partial_\mu \nu \partial^\mu \beta)^2+\sin^2 \theta (\partial_\mu \beta \partial^\mu \beta)\left(\partial^\mu \beta(2 \partial_\mu \nu +\partial_\mu \beta)\right)}}{\partial_\mu \beta \partial^\mu \beta}
\label{albenu}
\enr
In terms of which we define a four dimensional Clebsch field:
\beq
v_{C}^{\mu} \equiv \alpha \partial^\mu \beta + \partial^\mu \nu
\label{Clebsch}
\enq
Plugging \ern{Clebsch} and using the definitions of \ern{albenu} we obtain after some cumbersome but straight forward calculations the result:
\beq
{\cal L}_{KGc} = \bar \rho \left[v_{C}^{\mu} v_{C\mu} - c^2\right].
\label{LKGc2}
\enq
Defining the mass density in the rest frame as:
\beq
\rho_0 = \frac{\bar \rho}{c} \left[\sqrt{v_{C}^{\mu} v_{C\mu}} + c\right].
\label{LKGc3}
\enq
It follows that:
\beq
{\cal L}_{KGc} = c \rho_0 \left[\sqrt{v_{C}^{\mu} v_{C\mu}} - c\right] ={\cal L}_{{\rm Relativistic~Flow}} .
\label{LKGc4}
\enq
Thus the classical part of ${\cal L}_{KG}$ is equivalent (although in a non trivial way) to
a Lagrangian density of a classical relativistic fluid but of course without an internal energy
(see equation (103) in \cite{FluRel}). We note that unlike the non-relativistic Pauli spin flow which has a classical redundant term (in the sense that it does not comply with the fluid frame work) of the form (see equation (63) in \cite{Fisherspin3}):
\beq
\lim_{\hbar\rightarrow 0} \varepsilon_{qs} = \frac{1}{2} (1-\alpha^2) (\vec \nabla \beta)^2
\enq
In Dirac theory we have a perfect mapping between the classical parts of Dirac's Lagrangian and the relativistic flows, without any "left overs".

\section {The Dirac quantum term}

Let us now compare the Dirac quantum term ${\cal L}_{KGq}$ appearing in \ern{LKG3} to the quantum Fisher information term appearing in equation (113) of \cite{FluRel}
which we cite here for completeness:
\beq
{\cal L}_{RFq} = \frac{\hbar^2}{2 m} \partial^\mu a_0 \partial_\mu a_0, \qquad a_0 \equiv \sqrt{\frac{\rho_0}{m}}.
\label{Lagquantum2}
\enq
On a superficial consideration they look quite the same, however, looking more closely striking differences appear. First ${\cal L}_{KGq}$ depends on two "density amplitudes" (one for each spin) as opposed to the single amplitude of ${\cal L}_{RFq}$. Indeed, it is known that each energy eigenstate of the Dirac equation can accommodate two electrons each with a different spin. Second
a factor of $2$ is missing in the denominator of  ${\cal L}_{KGq}$. We shall try to answer the questions as follows, looking at \ern{LKGc3} we recall that the amplitudes $R_\uparrow$ and $R_\downarrow$ are not simply connected to the density as:
\beq
\rho_0 = \bar \rho \left[\sqrt{\frac{v_{C}^{\mu} v_{C\mu}}{c^2}} + 1\right].
\label{LKGc3b}
\enq
However, according to equation (104) of \cite{FluRel}:
\beq
\sqrt{v_{C \mu} v_{C}^{\mu}} = |v_{C 0}|\sqrt{1-\frac{\vec v_{C}^2}{v_{C 0}^2}}
= |v_{C 0}|\sqrt{1-\frac{\vec v^2}{c^2}} =\frac{|v_{C 0}|}{\gamma}
\label{Lagactionsimpb5}
\enq
Also according to equation (101) of \cite{FluRel}:
\beq
|v_{C 0}|= c \lambda
\label{Lagactionsimpb6}
\enq
For a classical fluid lacking internal energy and satisfying the equations of motion (see equation (58) of \cite{FluRel}):
\beq
\lambda =\gamma
\label{Lagactionsimpb7}
\enq
thus up to quantum corrections:
\beq
\sqrt{v_{C \mu} v_{C}^{\mu}} \approx c
\label{Lagactionsimpb8}
\enq
Inserting \ern{Lagactionsimpb8} into \ern{LKGc3b}
\beq
\rho_0 \approx 2 \bar \rho.
\label{rho0}
\enq
Thus:
\beq
a_0 = \sqrt{\frac{\rho_0}{m}} \approx  \sqrt{\frac{2 \bar \rho}{m}} = \sqrt{2} R,
\qquad R^2 \equiv R_\downarrow^2 + R_\uparrow^2
\label{a0}
\enq
In terms of $R$ and $\theta$ one may write the quantum part of the Lagrangian density as:
\beq
{\cal L}_{KGq} = \frac{\hbar^2}{m} \left(\partial^\mu  R_\uparrow  \partial_\mu
R_\uparrow +\partial^\mu  R_\downarrow  \partial_\mu R_\downarrow\right)
= \frac{\hbar^2}{m} \left(\partial^\mu  R  \partial_\mu
R +R^2 \partial^\mu  \theta  \partial_\mu \theta\right).
\label{KGq2}
\enq
Thus:
\beq
{\cal L}_{KGq} \approx \frac{\hbar^2}{2m} \left(\partial^\mu  a_0  \partial_\mu
a_0 +a_0^2 \partial^\mu  \theta  \partial_\mu \theta\right).
\label{KGq2c}
\enq
We notice that in Dirac's theory $R$ is not a probability amplitude as according to
\ern{Jprob2}
\beq
J^0 = \psi_1^\dagger \psi_1 + \psi_2^\dagger \psi_2  = R^2 +  \psi_2^\dagger \psi_2 \ge R^2.
\label{Jprob3}
\enq
thus the second term in the quantum Lagrangian density perhaps is not surprising, and of course
a complete calculation requires the inclusion of the quantum effects neglected in \ern{Lagactionsimpb8}.

\section {Conclusion}

We have shown the equivalence of the classic sector of Dirac theory to relativistic fluid dynamics.
This solves the riddle about some strange terms appearing in the fluid description of Pauli's theory. However, the quantum sector of Dirac's theory contains an additional term (the same "redundant" terms appears also in the fluid representation of Pauli's theory \cite{Fisherspin3}), which is not expected based on purely Fisher information considerations. Thus a deeper study is warranted, taking into account both quantum contributions to the $\lambda$ term which is a property of the relativistic fluid and also the unique definition of the probability density in Dirac's theory taking into account all four spinor amplitudes. This important task is left for the future.

Of course a complete description of the physics of the electron in terms of Dirac theory will require the interaction of the electron with the electromagnetic field which imply four potential terms in the variational action. This also is left for future endeavours.

Finally we remark that the nature of the quantum relativistic flow remains quite mysterious. One cannot avoid the obvious question: "a flow of what?" This fundamental question has consequences for both the issues raised above (the strange Fisher information addition and the existence of electromagnetic fields). We offer a rather bold hypothesis that dates back to Riemann regarding the geometry of space time. According to Riemann \cite{Weyl} all physical entities are geometrical, hence the flow is just the geometrical description of some thin elongated defect in space-time (thin in spatial dimensions but elongated in the temporal direction), the position of this defect is not well defined which is the reason for the appearance of the Fisher information term. We recall, that based on Riemann's proposal Einstein suggested the very successful theory of general relativity \cite{Einstein} describing gravity to high precision as the metric of space-time. However, an attempt of Weyl \cite{Weyl} to geometrize the electromagnetic field based on affine geometry is regarded as less successful. Also the idea of Schr\"{o}dinger \cite{Schrodinger} to geometrize matter based on the non-symmetric affine connection is not considered successful. Yet we are hopeful that the current mapping of relativistic flow to Dirac theory may shed some light on those early attempts and some progress can be made.

\end {document}